\documentclass[aps,pra,twocolumn,showpacs,1-pt]{revtex4-1}
\usepackage{ulem}
\usepackage{amsmath}
\usepackage{amsfonts}
\usepackage{amssymb}
\usepackage{fancyhdr}
\usepackage{color}
\usepackage{epsfig}
\usepackage{epstopdf}

\renewcommand{\[}{\begin{eqnarray}}
\renewcommand{\]}{\end{eqnarray}}
\newcommand{\ket}[1]{ | #1 \rangle}
\newcommand{\bra}[1]{ \langle #1 | }

\begin{document}

\title{Clearer visibility Hong-Ou-Mandel effect with correlation function based on rates rather than intensities}

\author{Krzysztof Roso\l{}ek$^1$, Kamil Kostrzewa$^1$, Arijit Dutta$^{1,2}$, Wies{\l}aw Laskowski$^1$, Marcin Wie\'sniak$^{3}$, Marek \.Zukowski$^1$}

\address{$^1$ Institute of Theoretical Physics and Astrophysics, Faculty of Mathematics, Physics and Informatics, University of Gda\'{n}sk, 80-308 Gda\'{n}sk, Poland}
\address{$^2$ Department of Physics, Hanyang University, Seoul 04763, Korea}
\address{$^3$ Institute of Informatics, Faculty of Mathematics, Physics and Informatics, University of Gda\'{n}sk, 80-308 Gda\'{n}sk, Poland}

\begin{abstract}
We test ideas put forward e.g. in  arXiv:1508.02368, which suggest that using rates in quantum optics can lead to  better indicators of non-classicality for states of quantum optical fields with undefined photon numbers. By rate we mean the ratio of registered photons in a given detector to the total number of detected photons in all detectors in a run of the experiment. 
For the Hong-Ou-Mandel effect for parametric down conversion fields,  we show that by using two detector correlation functions which are defined in terms of averages of products  of measured  rates, rather than usual intensities, one can observe non-classical visibilities beyond $1/2$ for significantly  higher pump rates. At such rates we have already have the partially stimulated emission, which leads to significant amplitudes for multiple pairs production, still the new approach allows to clearly see the non-classical dip. We consider both the perfect and imperfect detection efficiency. 
  
\end{abstract}

\maketitle

{\it Introduction.}
Quantum interference of higher order cannot be understood as an effect of superposed waves. While a repetition of single photon experiments, leads to statistics which is similar to the  predicted pattern of intensities for  Maxwellian electromagnetic waves, even the simplest two-photon situations, like the ones due to entanglement of pairs of photons lead to completely non-classical interference effects. Non-classical interference of two photons can occur also in situations which are not inspired by Bell's theorem. The most elementary effect of this kind is the Hong-Ou-Mandel dip ~\cite{HONG}, which relies entirely on the bosonic nature of photons, and is only due to their indistinguishability.

We shall use this phenomenon to illustrate our thesis, that in some situations redefined correlation functions better reveal the non-classical nature of some states of light. The redefinition is to replace the usual intensity correlation functions $G'(a; b)=\frac{\langle I_aI_b\rangle}{\langle I_a\rangle\langle I_b\rangle}$, or  $G(a; b)={\langle I_aI_b\rangle}$,
by $C(a; b)={\langle \frac{I_aI_b}{(I_a+I_b)^2}\rangle}$, where $I_x$ is the total intensity measured in beam $x=a,b$. This approach is inspired by Ref. \cite{ZukStokes}.
 
 The basic idea of the standard Hong-Ou-Mandel-type experiment is as follows.
Let us consider two photons entering  a $50:50$ beam splitter, one  via input $1$ and one via $2$. Let us assume that the photons are of the same polarization, and their wave packet profiles are identical, so
 that behind the beam splitter there is no way to tell which photon originates from which input port. If we imagine a symmetric $50:50$ beam splitter which upon reflection imposes a phase shift of $\pi/2$, the amplitudes of the two indistinguishable process, of two transmissions and of two reflections, differ by a phase shift $\pi$, and cancel out. Photons emerge only in pairs, via any of the two outputs, with equal probability, $1/2$.
 Two indistinguishable bosons like to go together, see Feynman's Lectures. 
 This photon bunching effect is called dip because if we make photons distinguishable, e.g. by a time delay before the beam splitter, beyond their coherence time, the situation is back to `normal'. We have equal probabilities for the photons to emerge in one of the exits, or via two exits. Therefore, if the temporal resolution of the detectors is 
 (much) less sharp then the coherence time of the photons, we register coincidences, while in the prefect indistinguishability case they are impossible. Thus, when changing the delay time we observe a dip in coincidences down to zero, which occurs when we have indistinguishability. This is accompanied with no interference effects at the single detector level.

 There is no way to explain the depth of the dip via the classical theory of light. The classical theory allows no more than $50\%$ drop (dip) of the coincidences.
We shall re-derive this below. This will be done for incoherent stochastic mixtures of pulsed light. We present a re-derivation of this well known result, because further on it will allow  make  straightforward the introduction of the modified description.

Imagine a $50 : 50$ beam splitter at which a stochastic mixture of  pulsed light impinges via its two input ports $1$ and $2$. Let us use the analytic signal approach to make the description simpler. The input to port $1$ is some pulse $E_1(t;\lambda) e^{i\phi_1(\lambda)}$ and the input to port $2$ is  $E_2(t;\lambda) e^{i\phi_{2}}(\lambda)$, where $\phi_{l}(\lambda)$, $l=1,2$, are mutually independent overall random phases. The pairs of pulses appear with a probability  $p(\lambda)$. The total  randomness of the relative phase is taken into account by assuming that $\langle e^{i[\phi_1(\lambda)\pm\phi_2(\lambda)]} \rangle_p= \sum_{\lambda}p(\lambda) e^{i[\phi_1(\lambda)\pm\phi_2(\lambda)]}=0$, i.e. every value of the phases has equal probability. This condition precludes  an appearance of any stable interference pattern at a single detector. If one introduces a   controlled phase shift $\Phi$ in front of one of the input ports, one sees no   dependence  on $\Phi$ of the averaged intensity at each of the exit ports (which is the other feature of the Hong-Ou-Mandel phenomenon). However interference effects are still observable, under some conditions, when one measures correlation functions of the intensities at the two exit ports.

For a given $\lambda$, the output intensity at the exit $a$ reads $I_a(t;\lambda)=\frac{1}{\sqrt{2}}|E_1(t;\lambda)e^{i\phi_1(\lambda)}+E_2(t;\lambda)e^{i\phi_2(\lambda)}|^2$, whereas at the other one it is
 $I_b(t';\lambda)=\frac{1}{\sqrt{2}}|E_1(t';\lambda)e^{i\phi_1(\lambda)}-E_2(t';\lambda)e^{i\phi_2(\lambda)}|^2.$ Therefore,  one has 
\[&I_a(t;\lambda)= \frac{1}{2}[I_1(t;\lambda)+I_2(t;\lambda)]&\\
&+ [E_1(t;\lambda)E^*_2(t;\lambda)e^{i[\phi_1(\lambda)-\phi_2(\lambda)]}+c.c],\nonumber &\] 
and 
\[&I_b(t';\lambda)= \frac{1}{2}[I_1(t';\lambda)+I_2(t';\lambda)]&\nonumber\\
&- [E_1(t';\lambda)E^*_2(t';\lambda)e^{i[\phi_1(\lambda)-\phi_2(\lambda)]
}+c.c].&\]

 Assume additionally, again only for the sake of simplicity, that the temporal length of the pulses is by  orders of magnitude shorter  than the integration time (time resolution) of the detectors of light, and the output of the detectors is proportional to the integrated intensity.
If we drop the terms which on average vanish due to the relative phase randomness,
the product of the  time integrated intensities reads
for each $\lambda$
\begin{eqnarray}\label{CLASSICAL}
& I_{a}(\lambda)I_b(\lambda)  =  \int dt \int dt' I_{a}(t,\lambda)I_b(t', \lambda) 
&\nonumber \\
&=\frac{1}{4}\big[(I_1(\lambda)+I_2(\lambda))^2&\nonumber\\& - 2\big|\int dt E_1(t;\lambda)E^*(t;\lambda))\big|^2\big],
&
\end{eqnarray}
where $I_l(\lambda)=\int dt |E_l(t, \lambda)|^2.$
The maximal value of the negative term is, by the Cauchy inequality, bounded  by $2I_1(\lambda)I_2(\lambda)$. As $\sqrt{I_1(\lambda)I_2(\lambda)}\leq\frac{1}{2}(I_1(\lambda)+I_2(\lambda))$, this in turn means that
\begin{eqnarray}\label{CLASSICAL-2}
& I_{a}(\lambda)I_b(\lambda) \geq
\frac{1}{8}(I_1(\lambda)+I_2(\lambda))^2.
&
\end{eqnarray}

 The lowest possible value of $ I_{a}(\lambda)I_b(\lambda) $, in relation to the term $\frac{1}{4}(I_1(\lambda)+I_2(\lambda))^2$,  occurs when  $E_1(t;\lambda)=E_2(t;\lambda)$. In such case the positive term in (\ref{CLASSICAL}) is two times greater than the negative one. 

The highest is when $\int dt E_1(t;\lambda)E_2^*(t;\lambda))=0, $ that is for orthogonal, in the sense of the Hilbert space, analytic signals. In such a case we have only the positive term and $\big(I_{a}(\lambda)I_b(\lambda)\big)_{max}=\frac{1}{4}(I_1(\lambda)+I_2(\lambda))^2$.

Such an orthogonality can be achieved e.g.  if one introduces such a temporal delay between the pulses in beam $1$ and $2$, so that they never overlap temporarily (but the delay is much lower than the time resolution of the detection). For a more elaborate description involving classical electromagnetic fields, the orthogonality can be also due to e.g. perpendicular linear polarizations. 
  
Upon averaging of (\ref{CLASSICAL}) over the distribution $p(\lambda)$, we get 
\begin{equation}
G=\langle  I_{a}(\lambda)I_b(\lambda) \rangle_p.
\end{equation}

 As relation (\ref{CLASSICAL-2}) holds, the maximal depth of the dip, when one e.g. changes the temporal overlap of the pulses in the input beams, satisfies 
\begin{equation}
V_{class}=\frac{G_{max}-G_{min}}{G_{max}} \leq \frac{1}{2}.
\end{equation}

Breaching the value of $\frac{1}{2}$
indicates the non-classical nature of the observed light. The first demonstrations of the Hong-Ou-Mandel dip we done for very low intensities.
The original paper described experiments in which the two photons were originating from a spontaneous down conversion event (type I). 
Later on an experiment with even weaker detection rates showed violation of the classical bound by photons originating from independent sources \cite{KALTENBAEK}.

Recently experimenters  searching  non-classicality moved also to the stronger pumping regime. E.g.,
 in~\cite{Masza} authors consider a (very) bright squeezed vacuum state generated by a single source via frequency-degenerate type-II parametric down conversion (PDC). Bright (nearly $10^{6}$ photons per mode on average) twin beams interfere on a balanced beam splitter. Instead standard coincidence function the observable chosen to measure the effect was the variance of the photon number difference at the outputs normalized by the average of the total number of photons, thus the results are not directly related with the analysis to be presented here.

 We shall show that for an  intermediate pumping the modified correlation functions allow one to see clearly strictly quantum interference of visibility beyond $50\%$. 
We present an approach to data analysis  for the HOM interference measurement,  based on the tricks introduced in~\cite{ZukBell, ZukStokes}. In~\cite{ZukBell} the authors present a new approach to Bell inequalities for quantum optical fields. The new concept is to replace averages of intensities by averages of rates i.e. the normalized intensities. It leads to removal of some loopholes inherent in earlier approaches to quantum optical Bell inequalities~\cite{Reid}. In~\cite{ZukStokes} the same authors introduce some modification of the standard Stokes parameters. The modified Stokes operators are normalized and the vacuum component is removed. The replacement is as follows
\[
S_{\theta } = a_{\theta }^{\dagger}a_{\theta } - a_{\theta^{\perp}}^{\dagger}a_{\theta^{\perp}} \rightarrow \hat{\Pi}_{NV}\frac{a_{\theta }^{\dagger}a_{\theta } - a_{\theta^{\perp}}^{\dagger}a_{\theta^{\perp}}}{\hat{N}}\hat{\Pi}_{NV}, \label{newStokes}
\]
where $\hat{\Pi}_{NV}$ is the projection operator on non-vacuum sectors i.e. $\hat{\Pi}_{NV} = \hat{1} - \ket{\Omega}\bra{\Omega}$ and $\theta, \theta^{\perp}$ denote any pair two orthogonal polarizations (generally elliptic), while in  the denominator we have the operator for the total number of photons $\hat{N}=a_{\theta }^{\dagger}a_{\theta } + a_{\theta^{\perp}}^{\dagger}a_{\theta^{\perp}}$, which is independent of $\theta$. 
This new definition allows to construct improved entanglement criteria for multi-photon states. 


The state of light which we use here as our working example is the two-mode squeezed vacuum. This is because such is the state emitted by the parametric down conversion process when we crank up the pumping. Thus, the results can be readily tested in a laboratory.


\indent  First  we discuss the case of classical fields. In the second section we move to describing the  standard quantum correlation functions and the redefined ones. 

\section{Classical case and new correlation measure}
In the classical case, the redefined correlation function, for the same physical conditions as discussed in Introduction, reads 
$$C=\left\langle{\frac{\int dtI_a(t)\int dtI'_b(t')}{I_{tot}^2}}\right\rangle_p,$$
where $I_{tot}=\int dtI_a(t)+\int dI_b(t).$ 
Note that formally the formula for $C$ can be looked at as the one for $G$ in which 
 the input field amplitudes  $E_1(t)$, and $E_2(t)$ are replaced by $E_1(t)/I_{tot}$, and $E_2(t)/I_{tot}$, respectively. Therefore as  $V_{max}=50\%$ is the maximum possible  visibility for $G$ for any inputs, the maximum visibility for $C$ cannot breach this value. We have $$V'=\frac{C_{max} - C_{min}}{C_{max}}\leq \frac{1}{2}.$$

\section{Quantum case}
Let us define two following quantum correlation functions
\[
G_{Q}=\frac{{Tr [\varrho n_{a} n_{b}}]}{{Tr[ \varrho (n_{a} + n_{b})^2}]}\label{G_Q},
\]
and
\\
\[
C_{Q}=Tr[ \varrho \hat{\Pi}_{NV}\frac{n_{a} n_{b}}{(n_{a}+n_{b})^2}\hat{\Pi}_{NV} ]\label{G2},
\] 
where $n_a$ and $n_b$ are photon number operators in the exit modes of the beam splitter  a and b, respectively, and $\varrho$ represents the state of the quantum field. Function $G_{Q}$ follows the traditional approach and is a quantum version of $G$.  The denominator used here is only for the sake of a direct comparison with $C_Q$. It cancels out from the definition of the visibility.   The new approach is represented by function $C_Q$. 
Note that the observable in formula for the correlation function $C_{Q}$ can be rewritten as
\[
\hat{\Pi}_{NV}\frac{n_{a}}{\hat{N}}\frac{n_{b}}{\hat{N}}\hat{\Pi}_{NV},
\]
where $\hat{N}$ is the operator of the total number of photons. We have a product of the observed rates at the two channels, understood as 
ratios of the number of photons in the given channel to the total number of detected  photons.  Thus, the  definition is fully concurrent with the approach of~\cite{ZukBell,ZukStokes}, and formula (\ref{newStokes}).

\subsection{Detection of non-classicality in Hong-Ou-Mandel interference. }
Let us compare the two quantum correlation functions, using our example of Hong-Ou-Mandel interference.
As the input to the beam splitter we shall use our working example, the two-mode squeeze vacuum which results in the case of strongly pumped 
parametric down conversion source. That is the same one as in the case of the original Hong-Ou-Mandel experiment, however of much higher intensity, so that emission of multiple pairs is possible. 



Let us start with the description of the  state which will be used in our calculations. 
The simplified interaction Hamiltonian for  a type-I frequency-degenerate PDC process is of the form
\[
\mathcal{H}=i\chi a_1^{\dagger }a_2^{\dagger } + h.c, 
\]
where $a_1$ and $a_2$ are annihilation operators for photons to be fed into the two input modes of the beam splitter,  $\chi$ is a coupling constant depending on the nonlinearity of the crystal and the power of the pumping field. If we take the vacuum as an initial state of the field  we get a two-mode bright squeezed vacuum 
\[
\ket{BSV}=\frac{1}{\cosh{\Gamma}}\sum\limits_{n=0}^{\infty}\tanh^{n}{\Gamma}\ket{n_{(1)},n_{(2)}}, \label{BSV}
\]
where $\Gamma$ is a gain parameter, and $\ket{n_{(1)},n_{(2)}}$ denotes $ \frac{1}{n!}  a_1^{\dagger n} 
a_2^{\dagger n} \ket{\Omega}$, where in turn $\Omega$ denotes photon vacuum in all modes. Each component of this state is composed of $2n$ photons, distributed equally in spatial modes 1 and 2 ($n$ in each mode). If the phonons in modes $a_1$ and $a_2$ have identical wave packet profiles and polarizations, the state $\ket{BSV}$ describes  the situation which warrants the perfect indistingushability of the photons behind the beam splitter. We shall assume that the beam splitter's action is described in this case by   $a_1 \rightarrow \frac
{1}{\sqrt{2}}(a+b)$,   $a_2 \rightarrow \frac
{1}{\sqrt{2}}(a - b)$, where $a$ and $b$ describe the annihilation operators in output modes which bear the same names. Such desciption is in full concurrence with the classical operation of the beam splitter introduced  earlier.

In the case of partial distiguishability we must make our description more refined. We introduce also creation (and annihilation) operators for the photons which are distinguishable for the ones created by $a_1^\dagger$ and $a_2^\dagger$, let us denote them by $a_{1\perp}^\dagger$ and 
$a_{2\perp}^\dagger$. The subscripts $1,2$ as before denote the beams, while $\perp$ denote the fact that these operators represent (some) orthogonal modes, be it in wave packet profiles, or polarizations. This technically means that $[a_{r},a_{s\perp}]=[a_{r},a_{s\perp}^\dagger]=0$, where $r,s=1,2$.

Assume that in one of the beams, say 2, we introduce a time delay, or polarization rotation,  which results in photons which are partially distinguishable behind the beam splitter with the ones of beam 1. This can be described via a replacement of $a_2^\dagger$ by $\cos{\alpha} a_2^\dagger+\sin{\alpha}a_{2\perp}^\dagger$ in the formula  (\ref{BSV}), where $\alpha$ paramatrizes the degree of distiguishability.
We get  a modified state
\[
\ket{BSV}_{\alpha } = \frac{1}{\cosh{\Gamma}} \sum_{n=0}^{\infty} \frac{1}{n!} \tanh^{n}\Gamma a_1^{\dagger n} \label{BSValfa} \\ \nonumber
\times(\cos{\alpha} ~a_2^{\dagger}+\sin{\alpha} ~a_{2\perp}^{\dagger})^n \ket{\Omega}. 
\] Notice that the mode $a_{1\perp}$ is unoccupied.

The (relevant) beam splitter transformations acting on the input state  are as follows: $a_1 \rightarrow \frac
{1}{\sqrt{2}}(a+b)$,   $a_1 \rightarrow \frac
{1}{\sqrt{2}}(a-b)$, and $a_{2\perp} \rightarrow \frac
{1}{\sqrt{2}}(a_{\perp}-b_{\perp})$, where $a_{\perp},b_{\perp}$ are annihilation operators for the exit modes of  the beam splitter which carry the photons which were in the mode $a_{2\perp}$ of the input beam $2$. Obviously $[a,a_\perp]=[a,a_\perp^\dagger]=0$
and  $[b,b_\perp]=[b,b_\perp^\dagger]=0$, as well as $[a,b_\perp]=[a,b_\perp^\dagger]=0$.

After all replacements  corresponding to the beam splitter transformation the  state $\ket{BSV}_\alpha$ takes the form (see Appendix)
\begin{eqnarray}
\mathcal{U}_{BS}\ket{BSV}_{\alpha}  \nonumber
& =& \sum_{n=0}^{\infty}\sum_{k=0}^{n}\sum_{l=0}^{n}\sum_{m=0}^{n-l}\sum_{p=0}^{l}B_{n}^{k,l,m,p} \nonumber \\ 
&\times& \sqrt{(2n-k-l-m)!(l-p)!k!p!}\\
&\times& \ket{2n-k-l-m_{(a)},l-p_{(a_\perp)},k_{(b)},p_{(b_\perp)}}, \nonumber
\label{BSVout}
\end{eqnarray}
where \[&B_{n}^{k,l,m,p}&\nonumber \\& = \frac{(-1)^{m+p}\tanh^{n}\Gamma\sin^l{\alpha}\cos^{n-l}{\alpha}}{n!2^n\cosh{\Gamma}}\binom{n}{k}\binom{n}{l}\binom{n-l}{m}\binom{l}{p},&
\] 
and \[\ket{j_{(a)},k_{(a_\perp)},l_{(b)},m_{(b_\perp)}} =\frac{1}{\sqrt{j!k!l!m!}}a^{\dagger j}a_\perp^{\dagger k}b^{\dagger l}b_\perp^{\dagger m}\ket{\Omega}, \nonumber \] i.e. letters $j,k,l,m$ represent numbers of photons in modes $a,a_\perp,b,b_\perp$.


For the two special cases, namely $\alpha = 0$ and $\alpha = \frac{\pi }{2}$,  that is full indistinguishability and full distinguishability, respectively,
one can easily  perform the full analytical  calculation (without cutting the sum over $n$). 

As was pointed out for $\alpha =0$, the full indistinguishability,  the initial state (\ref{BSValfa}) is simply $\ket{BSV}$. The beam splitter transforms it into following state
\begin{eqnarray}
\mathcal{U}_{BS}\ket{BSV}_{\alpha = 0} = \sum^{\infty}_{n=0}\sum^{n}_{k=0}A_{n}\binom{n}{k}(-1)^k \nonumber \\ 
\sqrt{(2k!)(2(n-k))!} \ket{2k_{(a)},2(n-k)_{(b)}},
\end{eqnarray}
where $A_{n} = \frac{\tanh^{n}\Gamma}{n!2^n\cosh{\Gamma}}$. It leads to the analytic expressions for $G_{Q}$ and $C_{Q}$. Namely
\begin{equation}
G_{Q}(\alpha = 0) = \frac{1}{4} \sinh ^2 \Gamma \text{sech} {2 \Gamma} \label{G_Qan}
\end{equation}
and
\begin{eqnarray}
C_{Q}(\alpha = 0) &=& \frac{1}{8} (\tanh ^2 \Gamma\nonumber \\&+&\text{sech}^2 \Gamma \ln \left(\text{sech}^2 \Gamma\right)). \label{G2an}
\end{eqnarray}

For $\alpha = \frac{\pi }{2}$, full distinguishability, after beam splitter the state reads
\[
\mathcal{U}_{BS}\ket{BSV}_{\alpha = \frac{\pi }{2}} = \sum^{\infty}_{n=0}\sum^{n}_{k=0}\sum^{n}_{l=0}A_{n}\binom{n}{k}\binom{n}{l}(-1)^l \nonumber \\ \sqrt{k!l!(n-k)!(n-l)!}\ket{k_{(a)},l_{(a_\perp)},n-k_{(b)},n-l_{(b_\perp)}}.
\]
The functions $G_{Q}$ and $C_{Q}$ are equal to
\[
G_{Q}(\alpha = \frac{\pi }{2}) = \frac{1}{8} (2-\text{sech}2 \Gamma) \label{G_Qpi}
\]
and
\begin{eqnarray}
C_{Q}(\alpha = \frac{\pi }{2}) &=& \frac{1}{8} (2 \tanh ^2 \Gamma \label{G2pi} \\
&+&\text{sech}^2 \Gamma \ln (\text{sech}^2 \Gamma )). \nonumber 
\end{eqnarray}

 With all that we can calculate the visibilities of the HOM experiment for $\ket{BSV}_{\alpha}$ state. One can identify maximal number of coincidence as the value of $G$ functions for $\alpha = \frac{\pi}{2}$ and minimal value for $\alpha = 0$. Hence the visibility reads 
\begin{equation}
V_{f}(\Gamma) = \frac{f(\alpha = \frac{\pi }{2}) - f(\alpha = 0)}{f(\alpha = \frac{\pi }{2})},\label{def}
\end{equation}
where $f = G_{Q}, C_{Q}$. Using expressions (\ref{G_Qan}), (\ref{G2an}), (\ref{G_Qpi}), (\ref{G2pi}) one obtains
\begin{equation}
V_{G_{Q}}(\Gamma) = \frac{1}{2-\text{sech}2 \Gamma} \label{vis1}
\end{equation}
and 
\begin{equation}
V_{C_{Q}}(\Gamma) =  \frac{\sinh ^2 \Gamma}{\cosh 2 \Gamma+\ln \left(\text{sech}^2 \Gamma\right)-1}. \label{vis2}
\end{equation}

\begin{figure}[!h]
\includegraphics[width=0.47\textwidth]{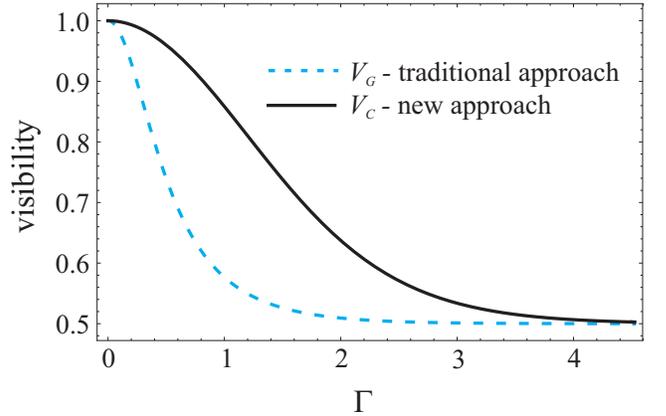} 
\caption{\label{vis} The visibilities versus parametric gain $\Gamma$.} 
\end{figure}
Fig.(\ref{vis}) shows the visibilities $V_{G_{Q}}$ and $V_{C_{Q}}$ with respect to the parametric gain $\Gamma$. Please notice that for all non-zero values of the  parametric gain $\Gamma$ the difference $\Delta V := V_{C_{Q}} - V_{G_{Q}}$ is positive, that is the modified definition of the correlation function beats the standard one in the visivility of the HOM dip. Both quantum visibilities are greater than the classical threshold $\frac{1}{2}$ and asymptotically tend to it.

\subsection{Imperfect detection efficiency}
In this section we consider imperfect ($\eta < 1$) detection efficiency. We assume that losses in each mode are independent and equal and model them by Bernoulli distribution with probability of success (photon registration) $\eta$ corresponding directly to the detection efficiency. All formulas which we use to calculate the visibilities are in the Appendix. The results are presented in Fig. \ref{vis_eff}.
 	
\begin{figure}[!h]
\includegraphics[width=0.47\textwidth]{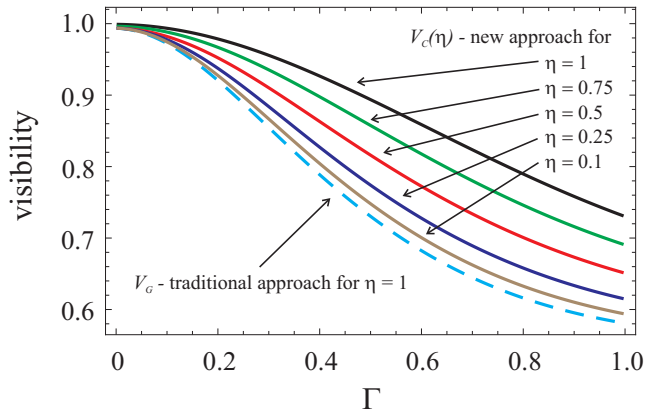} 
\caption{\label{vis_eff} The visibilities $V_{G_{Q}^{\eta}}$ and $V_{C_{Q}^{\eta}}$ with respect to the parametric gain $\Gamma$ for $\eta = 1$, $0.75$, $0.5$, $0.25$, $0.1$ for the new approach and for $\eta = 1$ for traditional one. Please notice that irrespectively of  $\eta$ the approach which is based on rates leads to the clearer visibility. Moreover, if $\eta$ is very small the new visibilities tend to traditional one with perfect efficiency.    } 
\end{figure}

\section{Conclusions}
The above considerations show that one can extend the applicability of the ideas of Ref. \cite{ZukStokes} to optical phenomena which are not polarization based. Also the effect which we study here is usually in not associated with entanglement  (at least at the two-photon level), thus we extend the approach beyond the entanglement detection  problems studied in 
\cite{MULTI}. 
The main difference with the usual approaches is that we divide the obtained intensities (here modeled by photon numbers) by the total registered intensity (all at a given run of the experiment). As we are interested in measurements of correlations this does not pose any additional difficulties in the experiments. To monitor correlations one must register coincident photon counts at the two outputs for each run of the experiment, and thus the total registered intensity is at hand, as the sum of the coincident intensities.

As the approach with rates leads to a clearer visibility of the Hong-Ou-Mandel effects, for higher pump values and this holds for quite low detection efficiencies, we hope that it will inspire further modifications of this type in the theory of correlation functions for quantum optics, moving further ahead modifications the standard methods, as those e.g. in \cite{GLAUBER}, \cite{MANDEL-WOLF}.

\acknowledgments 
The work is part of project BRISQ2 (2012-2015),  within EU FP7 (grant agreement No. 308803) and  was subsidized form funds for science for years 2012-2015 approved for international co-financed project BRISQ2 by Polish Ministry of Science and Higher Education (MNiSW). 
KK was supported by NCN Grant No. 2012/05/E/ST2/02352.
At a later stage  MZ was supported by an ERC grant QOLAPS, and a DFG-FNP award-grant COPERNICUS. 
WL acknowledges support by the EU grant ERC AdG QOLAPS. 
MW was supported by NCN Grant 2015/19/B/ST2/01999.
AD was supported by the National Research Foundation of Korea (NRF) grant funded by the Korea government (MSIP) (No. 2014R1A2A1A10050117).
We thank profs M. Chekhova and H. Weinfurter for discussions.


\appendix
\section{$50 :50$ Beam splitter transformation for $\ket{n_{(1)},n_{(2)}}$ state}

We can express $\ket{n_{(1)},n_{(2)}}$, which is two-mode $2n$-photon Fock state, in terms of creation operators formalism as
\[
\ket{n_{(1)},n_{(2)}} = \frac{1}{n!} (a_{1}^{\dagger} a_{2}^{\dagger})^n \ket{\Omega}.
\]
We denote the $50:50$ beam splitter transformation by $\mathcal{U}_{BS}$. Hence we get
\[
\ket{\psi } = \mathcal{U}_{BS}\ket{n_{(1)},n_{(2)}} = \\ \nonumber \frac{1}{2^n} \frac{1}{n!} [(a^{\dagger}+b^{\dagger})(a^{\dagger}-b^{\dagger})]^n \ket{\Omega} \\ \nonumber
= \frac{1}{2^n} \frac{1}{n!} (a^{\dagger 2}-b^{\dagger 2})^n \ket{\Omega} \\ \nonumber
= \frac{1}{2^n} \frac{1}{n!} \sum^n_{k=0} {{n}\choose{k}}(-1)^k {a}^{\dagger 2(n-k)} {b}^{\dagger 2k} \ket{\Omega},
\]
where $a^{\dagger}$ and $b^{\dagger}$ are creation operators for modes after beam splitter. Moreover the operators $a^{\dagger}+b^{\dagger}$ and $a^{\dagger}-b^{\dagger}$ commute. 
Thus we have
\[&
\ket{\psi } = \frac{1}{2^n} \frac{1}{n!} \sum^n_{k=0} {{n}\choose{k}}(-1)^k \sqrt{(2(n-k))!(2k)!} &\nonumber  \\ 
&\times\ket{2(n-k)_{(a)},2k_{(b)}}. &\label{AfterBS}
\]
Notice that there are only two components in the state (\ref{AfterBS}), which reveal no coincidences, namely for $k=0$ and $k=n$.

\section{$50 :50$ Beam splitter transformation for $\ket{BSV}_{\alpha }$ state with distinguishable photons}

We consider the state (\ref{BSValfa})
\[
\ket{BSV}_{\alpha } = \frac{1}{\cosh{\Gamma}} \sum_{n=0}^{\infty} \frac{1}{n!} \tanh^{n}\Gamma a_{1}^{\dagger n} \\ \nonumber
\times(\cos{\alpha}a_{2}^{\dagger}+\sin{\alpha}a_{2 \perp}^{\dagger})^n \ket{\Omega} \nonumber ,
\]
where $a_{2}$ and $a_{2 \perp}$ are orthogonal modes and $\alpha \in [0,\frac{\pi }{2}]$ is a parameter which introduces a kind of distinguishability measure in the spatial mode $a_{2}$. Please notice that for $\alpha = 0$ we reproduce the two-mode bright squeezed vacuum state. After beam splitter the state takes the following form 
\[
\mathcal{U}_{BS}\ket{BSV}_{\alpha } = \sum_{n=0}^{\infty}A_{n}(a^{\dagger}+b^{\dagger})^n \nonumber \\ \left(\cos{\alpha}(a^{\dagger}- b^{\dagger})+\sin{\alpha}(a_{\perp}^{\dagger}-b_{\perp}^{\dagger})\right)^n \ket{\Omega},
\]
where $A_{n} = \frac{\tanh^{n}\Gamma}{n!2^n\cosh{\Gamma}}$. Using the binomial expansion we get
\[
\mathcal{U}_{BS}\ket{BSV}_{\alpha } = \sum_{n=0}^{\infty}\sum_{k=0}^{n}\sum_{l=0}^{n}\sum_{m=0}^{n-l}\sum_{p=0}^{l}B_{n}^{k,l,m,p} \nonumber \\ (a^{\dagger})^{2n-k-l-m}(b^{\dagger})^{k+m}(a_{\perp}^{\dagger})^{l-p}(b_{\perp}^{\dagger})^{p}\ket{\Omega},
\]
where 
\[
&B_{n}^{k,l,m,p} = \frac{(-1)^{k+p}\tanh^{n}\Gamma\sin^l{\alpha}\cos^{n-l}{\alpha}}{n!2^n\cosh{\Gamma}}& \\ \nonumber
&\times\binom{n}{k}\binom{n}{l}\binom{n-l}{m}\binom{l}{p}. &
\]
Thus, as superposition of Fock states the exit state reads 
\[
\mathcal{U}_{BS}\ket{BSV}_{\alpha } = \sum_{n=0}^{\infty}\sum_{k=0}^{n}\sum_{l=0}^{n}\sum_{m=0}^{n-l}\sum_{p=0}^{l}B_{n}^{k,l,m,p} \label{BSValfaA} \\ \nonumber 
\times \sqrt{(2n-k-l-m)!(l-p)!k!p!} \\ \nonumber 
\times \ket{2n-k-l-m_{(a)},l-p_{(a_{\perp})},k_{(b)},p_{(b_{\perp})}} \nonumber .
\]

\section{Calculation of the expressions for $G_{Q}$ and $C_{Q}$}

We present all calculations concerning the expressions (\ref{G_Q}) and (\ref{G2}), i.e.
\[
G_{Q}=\frac{{Tr [\varrho n_{a} n_{b}}]}{{Tr[ \varrho (n_{a} + n_{b})^2}]} \label{GQA}
\] 
and
\[
C_{Q}=Tr[ \varrho \hat{\Pi}_{NV}\frac{n_{a} n_{b}}{(n_{a}+n_{b})^2}\hat{\Pi}_{NV}] \label{CQA} . 
\] 
In order to calculate expectation values (\ref{G_Q}) and (\ref{G2}) we use the fact that $f(\hat{n})\ket{n} = f(n)\ket{n}$, where $f$ is a certain function of a photon number operator.																												
To calculate expressions (\ref{GQA}) and (\ref{CQA}) for state (\ref{BSValfaA}) for any $\alpha$, we should take summation to infinity into account. For $\alpha \neq 0, \frac{\pi}{2}$ this is rather cumbersome  and requires a cut-off.  However, the only cases we consider in this paper are $\alpha = 0$ and $\alpha = \frac{\pi}{2}$. In case $\alpha = 0$ the state (\ref{BSValfaA}) reduces to

\begin{eqnarray}
\mathcal{U}_{BS}\ket{BSV}_{\alpha = 0} = \sum^{\infty}_{n=0}\sum^{n}_{k=0}A_{n}\binom{n}{k}(-1)^k \nonumber \\ 
\sqrt{(2k!)(2(n-k))!} \ket{2k_{(a)},2(n-k)_{(b)}},
\end{eqnarray}
where $A_{n} = \frac{\tanh^{n}\Gamma}{n!2^n\cosh{\Gamma}}$.
In accordance with definitions (\ref{GQA}) and (\ref{CQA})   using known properties of Fock states one obtains
\[
&G_{Q}(\alpha = 0)&\nonumber \\ &= \frac{\sum_{n}\sum_{k}A_{n}\binom{n}{k}(2k!)(2(n-k))!(4k(n-k))}{\sum_{n}\sum_{k}A_{n}\binom{n}{k}(2k!)(2(n-k))!(4n^2)} & \label{G0}
\]
and
\[&
C_{Q}(\alpha = 0) &\nonumber \\&= \sum_{n}\sum_{k}A_{n}^2\binom{n}{k}^2(2k!)(2(n-k))!\frac{k(n-k)}{n^2}.& \label{C0}
\]
All series in expressions (\ref{G0}) and (\ref{C0}) we can sum analytically. Therefore simplified formulas are as follows
\begin{equation}
G_{Q}(\alpha = 0) = \frac{1}{4} \sinh ^2 \Gamma \text{sech} {2 \Gamma} 
\end{equation}
and
\begin{eqnarray}
C_{Q}(\alpha = 0) &=& \frac{1}{8} (\tanh ^2 \Gamma\nonumber \\&+&\text{sech}^2 \Gamma \ln \left(\text{sech}^2 \Gamma\right)). 
\end{eqnarray}
Similarly we proceed in the case $\alpha = \frac{\pi}{2}$. The state (\ref{BSValfaA}) simplifies to
\[
\mathcal{U}_{BS}\ket{BSV}_{\alpha = \frac{\pi }{2}}  =& \\ \nonumber =\sum^{\infty}_{n=0}\sum^{n}_{k=0}\sum^{n}_{l=0}A_{n}\binom{n}{k}\binom{n}{l}(-1)^l & \nonumber \\ \times \sqrt{k!l!(n-k)!(n-l)!} \nonumber \\  \times \ket{k_{(a)},l_{(a_\perp)},n-k_{(b)},n-l_{(b_\perp)}}. \nonumber
\] 
Functions (\ref{GQA}) and (\ref{CQA}) take the form
\[
&G_{Q}(\alpha =\frac{\pi}{2}) = \\ \nonumber & =\frac{\sum_{n}\sum_{k}\sum_{l}A_{n}^2\binom{n}{k}\binom{n}{l}((k+l)(2n-k-l))}{\sum_{n}\sum_{k}\sum_{l}A_{n}^2\binom{n}{k}\binom{n}{l}(4n^2)}  \label{Gpi2}
\]
and
\[
& G_{Q}(\alpha = \frac{\pi}{2}) = \\ \nonumber & =\sum_{n}\sum_{k}\sum_{l}A_{n}^2\binom{n}{k}\binom{n}{l}\frac{(k+l)(2n-k-l)}{4n^2}.
\] \label{Cpi2}
Again all series in (\ref{Gpi2}) and (\ref{Gpi2}) summable and hence
\[
G_{Q}(\alpha = \frac{\pi }{2}) = \frac{1}{8} (2-\text{sech}2 \Gamma) 
\]
and
\begin{eqnarray}
C_{Q}(\alpha = \frac{\pi }{2}) &=& \frac{1}{8} (2 \tanh ^2 \Gamma  \\
&+&\text{sech}^2 \Gamma \ln (\text{sech}^2 \Gamma )). \nonumber 
\end{eqnarray}
Now, following the definition (\ref{def}) we get appropriate visibilities
\begin{equation}
V_{G_{Q}}(\Gamma) = \frac{1}{2-\text{sech}2 \Gamma} 
\end{equation}
and 
\begin{equation}
V_{C_{Q}}(\Gamma) =  \frac{\sinh ^2 \Gamma}{\cosh 2 \Gamma+\ln \left(\text{sech}^2 \Gamma\right)-1}. 
\end{equation}

\section{Calculation of the expressions $G_{Q}$ and $C_{Q}$ for imperfect detection efficiency}

In order to calculate the visibility we consider only two cases, i.e. $\alpha = 0$ and $\alpha = \frac{\pi}{2}$. For detection efficiency $\eta$ and $\alpha = 0$ the traditional correlation function reads
\[
G_{Q}^{\eta}(\alpha = 0) = \sum_{n = 0}^{\infty}\sum_{k = 0}^{n} \sum_{l = 0}^ {2k} \sum_{m = 0}^{2(n-k)} l m \\ Z_{nk}^2(c_{2k}^{l})^2(c_{2(n-k)}^{m})^2, \nonumber
\]
\\
where $Z_{nk} = A_{n}(-1)^k\binom{n}{k} \sqrt{(2 k)! (2 (n-k))!} $. The coefficient $c_{y}^{x} = \sqrt{\binom{y}{x}(1-\eta )^{y-x} \eta ^{x}}$ is the probability amplitude of detecting $x$ photons, assuming $y$ photons reaching the detector. Similarly, the new correlation function equals
\[
C_{Q}^{\eta}(\alpha = 0) = \sum_{n = 0}^{\infty}\sum_{k = 0}^{n} \sum_{l = 0}^ {2k} \sum_{m = 0}^{2(n-k)} \frac{l m}{(m+l)^2} \\ Z_{nk}^2(c_{2k}^{l})^2(c_{2(n-k)}^{m})^2. \nonumber
\]
For $\alpha = \frac{\pi}{2}$ we obtain
\[
G_{Q}^{\eta}(\alpha = \frac{\pi}{2}) = \sum_{n = 0}^{\infty}\sum_{k = 0}^{n} \sum_{l = 0}^ {n} \sum_{m = 0}^{k}\sum_{p = 0}^{l}\sum_{s = 0}^{n-k}\sum_{r = 0}^{n-l} \\ (m+p)(s+r) Z_{nkl}^2(c_{k}^{m})^2(c_{l}^{p})^2(c_{n-k}^{s})^2(c_{n-l}^{r})^2 \nonumber
\]
and
\[
C_{Q}^{\eta}(\alpha = \frac{\pi}{2}) = \sum_{n = 0}^{\infty}\sum_{k = 0}^{n} \sum_{l = 0}^ {n} \sum_{m = 0}^{k}\sum_{p = 0}^{l}\sum_{s = 0}^{n-k}\sum_{r = 0}^{n-l} \\ \frac{(m+p)(s+r)}{(m+p+s+r)^2} Z_{nkl}^2(c_{k}^{m})^2(c_{l}^{p})^2(c_{n-k}^{s})^2(c_{n-l}^{r})^2, \nonumber
\]
\\
where $Z_{nkl} = A_{n}(-1)^l \binom{n}{k} \binom{n}{l} \sqrt{k! l! (n-k)! (n-l)!}$. \\ 

We cut all sums to the certain $N_{max}$. For instance for $\Gamma < 1$, $N_{max} = 8$ seems to be good enough. Obviously, all formulas with $\eta$ reduce to the previous ones for $\eta = 1$ (perfect efficiency).  
\end{document}